\documentclass[12pt]{iopart}
\usepackage{iopams}  
\usepackage{graphicx}
\begin{document}

\title{Review of Conformally Flat Approximation for Binary Neutron Star Initial Conditions}

\author{In-Saeng Suh}

\address{Center for Astrophysics, Department of Physics and Center for Research Computing,
University of Notre Dame, Notre Dame, Indiana 46556, USA}
\ead{isuh@nd.edu}

\author{ Grant J. Mathews and J. Reese Haywood}
\address{Center for Astrophysics, Department of Physics, University of Notre Dame, Notre Dame, Indiana 46556, USA}
\ead{gmathews@nd.edu}
\ead{reese.haywood@gmail.com}

\author{N. Q. Lan}
\address{Hanoi National University of Education, 136 Xuan Thuy, Hanoi, Vietnam
and
Joint Institute for Nuclear Astrophysics (JINA), University of Notre Dame, Notre Dame, Indiana 46556, USA}
\ead{nquynhlan@hnue.edu.vn}

\vspace{10pt}
\begin{indented}
\item[]January 2016
\end{indented}

\begin{abstract}
The spatially conformally flat approximation  (CFA) is a viable method to deduce initial conditions
for the subsequent evolution of binary neutron stars employing the full Einstein equations.  Here we review the status of the original formulation of the CFA  for the general relativistic hydrodynamic 
initial conditions of binary neutron stars. 
We illustrate the stability of the conformally flat condition on the hydrodynamics 
by numerically evolving $\sim 100$  quasi-circular orbits. 
We illustrate  the use of this approximation for  orbiting neutron stars  in the quasi-circular orbit approximation to  demonstrate  the equation of state 
dependence of these initial conditions and how they might affect the emergent gravitational wave frequency as the stars approach the innermost stable circular orbit. 
\end{abstract}

%
\vspace{2pc}

\section{Introduction}

The epoch of gravitational wave astronomy has now begun with the first detection \cite{aligodetect}  of the merger of binary black holes by Advanced LIGO \cite{aLIGO}.
Now that the first ground based gravitational wave detection has been
achieved, observations of binary neutron star mergers should  soon be forthcoming.  This is particularly true  as 
 other  second generation observatories such as  Advanced  VIRGO \cite{VIRGO} and KAGRA \cite{KAGRA} will soon be online.  
In addition to binary black holes,  neutron-star 
binaries are thought to be among  the best candidate sources gravitational radiation \cite{ligoweb,thorne96}.
The number of such systems detectable by Advanced LIGO is estimated 
\cite{thorne96,Harry10,shaprev,bail96,tutu93,phin91,Kalogera04,Abadie10} to be  of order several events per year 
based upon observed close binary-pulsar systems \cite{burgay03,Lattimer12}.  There is a difference between neutron-star mergers and black hole mergers, however, in that neutron star 
mergers involve the complex evolution of the matter hydrodynamic equations in addition to the strong gravity field equations.  Hence, one must carefully consider both the hydrodynamic and
field evolution of these systems.

To date there have been numerous attempts to
calculate theoretical templates for gravitational waves  from compact binaries based upon numerical and/or analytic approaches (see for example
\cite{apos96,droz97,sath00,buo03,bose05,ott06,Ajith14,Pannarale15,Agathos15,Clark15}).  However, most approaches utilize a combination of
Post-Newtonian (PN)  techniques supplemented with quasi-circular orbit calculations and then applying full GR for only the last few orbits before disruption.
In this paper we review the status of the hydrodynamic evolution in the spatially conformally flat metric approximation (CFA) as a means to compute stable initial conditions 
 beyond the range of validity of the PN  regime, i.e  
near  the last stable orbits.
We establish the numerical stability of this approach based upon many orbit simulations  of  quasi-circular orbits. 
 We also illustrate the equation of state (EOS) dependence of the initial conditions and associated gravitational wave emission.

When binary neutron stars are well separated, the Post-Newtonian (PN) approximation is sufficiently
accurate \cite{allen05}. In the PN scheme, the stars are often  treated as point masses, either with or without
spin.  At third order, for example, it has been estimated \cite{lb3pn,Blanchet14,Mishra15} that the errors due to assuming the stars are point masses
is less than one orbital rotation \cite{lb3pn} over the $\sim 16,000$ cycles that pass through the LIGO detector
frequency band \cite{thorne96}.
Nevertheless, it has been noted in many works 
\cite{Agathos15,wm95,wmm96,mmw98,mw00,Baiotti10,Bose10,Read13,Maselli13,Bauswein15a,Bauswein15b,Fryer15,Dietrich15} that relativistic hydrodynamic
effects might be evident even at the separations ($\sim 10 - 100$ km) relevant to the LIGO window.  

Indeed,  the templates generated by PN approximations, unless carried out to fifth and sixth order \cite{lb3pn,Blanchet14},
may not be accurate unless the finite size and proper fluid motion of the stars is taken into account.  
In essence, the signal emitted during the last phases of 
inspiral depends on the finite size and EoS through  the  tidal deformation of the neutron stars and the cut-off frequency when tidal disruption occurs.

Numeric and analytic simulations \cite{duez03, Marronetti03, Miller04, Miller05, Miller07, Uryu06, Kiuchi09,Bernuzzi15, DePietri15} of binary neutron stars have explored the approach to
the innermost stable circular orbit (ISCO).  While these simulations represent some of the most accurate
to date, simulations generally follow the evolution for a handful of  orbits and are based upon initial conditions of quasi-circular orbits obtained in the conformally flat approximation.
Accurate templates of gravitational radiation require the ability to stably and reliable calculate the orbit initial conditions.
The CFA provides a means to obtain accurate initial conditions  near the ISCO.  

The spatially conformally flat approximation to GR was  first developed in detail in \cite{wmm96}. That  original formulation, however,  contained a mathematical error first pointed out by Flanagan \cite{Flanagan99} and subsequently corrected in \cite{mw00}.  This error in the solution to the shift vector led to a spurious NS crushing prior to merger. 
The formalism discussed below is for the corrected equations.    Here, we discuss  the hydrodynamic solutions as developed in \cite{wm95,wmm96,mmw98,mw00,wmbook,haywood}. This CFA formalism includes much of the nonlinearity inherent in GR and leads set of coupled, nonlinear, elliptic field equations that can be evolved stably. We also note that an alternative spectral method solution to the CFA configurations was developed by  \cite{Gourgoulhon01,Taniguchi01}, and  approaches beyond the CFA have also been proposed \cite{Uryu06}.  However, our purpose here is to review the viability of the hydrodynamic solution without the imposition of a Killing vector or special symmetry.  This approach is the most adaptable, for example, to general  initial conditions such as that of arbitrarily elliptical orbits and/or arbitrarily spinning neutron stars.

Here, we review the original CFA  approach and associated  general relativistic hydrodynamics formalism  developed
in \cite{wmm96,mw00,wmbook,haywood} and illustrate that it can produce stable initial conditions anywhere between the post-Newtonian to ISCO regimes.
We quantify  how long this method takes to converge to quasiequilibrium and demonstrate the stability by subsequently integrating  up to $\sim 100$ orbits for a binary neutron star system.  
We also analyze the EoS dependence of these quasi-circular initial orbits and show how these orbits can be used to make preliminary estimates of the gravitational wave signal for the initial conditions.  

This review is organized as follows.  In Section 2  the basic method is summarized and in Section 3 a number of code tests are performed
in the quasi-equilibrium circular orbit limit to demonstrate the stability of the technique.
The EoS dependence of the initial conditions and associated gravitational wave frequency and binding energy of various systems is analyzed in Section 4. Conclusions are presented in Section 5.

\section{Method}\label{sec:code}

\subsection{Field Equations}
\label{derconflat}
The solution of the field equations and hydrodynamic equations of motion were first  solved in three
spatial dimensions and explained in detail in the 1990's in
\cite{wm95,wmm96} and subsequently further reviewed in \cite{wmbook,mw97}. Here, we present a brief summary to introduce
the variables relevant to the present discussion.  

One starts with the slicing of
spacetime into the usual one-parameter family of hypersurfaces separated by
differential displacements in a time-like coordinate as defined in the (3+1) ADM
formalism \cite{adm,york79}.

In Cartesian $x, y, z$ isotropic  coordinates, proper distance is
expressed as
\begin{equation}
\label{conftrans}
ds^2 = -(\alpha^2 - \beta_i\beta^i) dt^2 + 2 \beta_i dx^i dt +
\phi^4\delta_{ij}dx^i dx^j
\end{equation}
where the lapse function $\alpha$ describes the differential lapse of proper
time between two hypersurfaces. The quantity  $\beta_i$ is the shift vector
denoting the shift in space-like coordinates between hypersurfaces. The
curvature of the metric of the 3-geometry is described by a position dependent
conformal factor $\phi^4$ times a flat-space Kronecker delta ($\gamma_{ij} = \phi^4 \delta_{ij}$).
This conformally flat condition (together with the maximal slicing gauge, $tr\{K_{ij}\} = 0$)
requires \cite{york79},
\begin{equation}
2\alpha K_{ij} = D_i \beta_j+D_j \beta_i - \frac{2}{3} \delta_{ij} D_k \beta^k
\label{detweiler}
\end{equation}
where $K_{ij}$ is the extrinsic curvature tensor and $D_i$ are 3-space covariant
derivatives.
This conformally flat condition on the metric provides a numerically valid
initial solution to the Einstein equations.  The vanishing of the Weyl tensor
for a stationary system in three spatial dimensions guarantees that a
conformally flat solution to the Einstein equations exists.

One consequence of this conformally-flat approximation to the three-metric is that the emission
of gravitational radiation is not explicitly evolved.
Nevertheless, one can extract the gravitational radiation signal and the back reaction
via a multipole expansion \cite{wmm96,thorne80}.  An application to the determination of the gravitational wave emission from the quasi-circular orbits computed here
is given in  \cite{Lan16}.
The advantage of this approximation is that conformal flatness stabilizes and
simplifies the solution to the field equations.

As a third gauge condition, one can choose separate coordinate transformations for the shift vector and the
hydrodynamic grid velocity to separately minimize the field and matter motion with respect to the coordinates.
This set of gauge conditions is key to the present application. It allows one to stably evolve up to hundreds and
even thousands of binary orbits without the numerical error associated with the frequent advecting of fluid through the grid.

Given a distribution of mass and momentum on some manifold, then one first solves the constraint equations
of general relativity at each time for a fixed distribution of matter. One then evolves the hydrodynamic equations
to the next time step. Thus, at each time slice a solution to the relativistic field equations and information on the
hydrodynamic evolution is obtained.

The solutions for the field variables $\phi$, $\alpha$, and $\beta^i$ reduce to simple Poisson-like equations
in flat space.
The Hamiltonian constraint \cite{york79}, is used to solve for the conformal factor $\phi$ \cite{wmm96,evansdiss},
\begin{equation}
\nabla^2{\phi} = -2\pi
\phi^5 \biggl[ W^2 (\rho(1 + \epsilon) + P) - P 
+ \frac{1}{16\pi} K_{ij}K^{ij}\biggr]~.
\label{phi}
\end{equation}
In the Newtonian limit, the RHS is dominated \cite{wmm96} by the proper matter
density $\rho$, but in strong fields and compact neutron stars there are also contributions from the
internal energy density $\epsilon$, pressure $P$, and extrinsic curvature.  The
source is also significantly enhanced by the  generalized curved-space Lorentz
factor $W$,
\begin{equation}
W = \alpha U^t~ = \biggl[ 1 + \frac{\sum{U_i^2}}{\phi^4}\biggr]^{1/2}~~,
\label{weq}
\end{equation}
where $U^t$ is the time component of the relativistic four velocity and $U_i$
are the covariant spatial  components. This  factor, $W$, becomes important near the last stable
orbit as the specific kinetic energy of the stars rapidly increases.

In a similar manner \cite{wmm96,evansdiss}, the Hamiltonian constraint,
together with the maximal slicing condition, provides an equation for the 
lapse function,
\begin{equation}
\nabla^2(\alpha\phi) = 2 \pi
\alpha \phi^5 \biggl[ 3W^2 [\rho (1 + \epsilon) + P] 
 - 2 \rho(1 + \epsilon) + 3 P
 + \frac{7}{16\pi} K_{ij}K^{ij}\biggr]~.
\label{alpha}
\end{equation}

Finally, the momentum constraints yields \cite{york79}
an elliptic equation for the shift vector \cite{mw00,flan99},
\begin{equation}
\nabla^2 \beta^i = \frac{\partial}{\partial x^i} \biggl(\frac{1}{3}
\nabla \cdot \beta \biggr) + 4 \pi \rho_3^i~,
\label{wilson3}
\end{equation}
where
\begin{equation}
\rho_3^i  = 4\alpha \phi^4 S_i + \frac{1}{4 \pi} \frac{\partial  ln(\alpha/\phi^6)}
{\partial x^j}
\biggl( \frac{\partial \beta^i}{\partial x^j} + \frac{\partial \beta^j}{\partial x^i}
-\frac{2}{3} \delta_{ij} \frac{\partial \beta^k}{\partial x^k} \biggr)~.
\end{equation}
Here the  $S_i$ are the spatial components of the momentum density one-form as defined
below.

We note that in early applications of this approach, the source for the shift vector contained a spurious term due to an incorrect transformation between contravariant and covariant forms of the momentum density as was pointed out in \cite{mw00,Flanagan99}.
As illustrated in those papers, this was the main reason why early hydrodynamic calculations induced a controversial  additional compression on stars causing 
them to collapse to black holes prior to inspiral \cite{wm95}.  This problem no longer exists in the formulation summarized here.

\subsection{Relativistic Hydrodynamics}
\label{hydro}

To solve for the fluid motion of the system in curved spacetime
it is convenient to use an Eulerian fluid description \cite{wilson79}.
One begins with the perfect fluid stress-energy tensor in the Eulerian observer rest frame,
\begin{equation}
T_{\mu\nu} = P g_{\mu \nu} + (\rho (1+ \epsilon) + P )U_\mu U_\nu ~,
\end{equation}
where $U_\nu$ is the relativistic four velocity one-form.

By introducing the usual  set of Lorentz contracted state variables it is possible to
write the relativistic hydrodynamic equations in a form
 which is reminiscent of their Newtonian counterparts \cite{wilson79}.
The hydrodynamic state variables are:
the coordinate baryon mass density,
\begin{equation}
D = W \rho~~;
\end{equation}
the coordinate internal energy density,
\begin{equation}
E = W \rho \epsilon~~;
\end{equation}
the spatial three velocity,
\begin{equation}
V^i = \alpha \frac{U_i}{\phi^4 W} - \beta^i~~;
\label{three-vel}
\end{equation}
and the covariant momentum density,
\begin{equation}
S_i = (D + E + PW) U_i~~.
\label{momeq}
\end{equation}

In terms of these state variables, the hydrodynamic equations in the CFA are as follows:
The equation for the conservation of baryon number takes the form,
\begin{equation}
\frac{\partial D}{\partial t}  =  -6D \frac{\partial \log\phi}{\partial t}
- \frac{1}{\phi^6} \frac{\partial}{\partial x^j}(\phi^6DV^j)~~.\
\end{equation}
The equation for internal energy evolution becomes,
\begin{eqnarray}
\frac{\partial E}{\partial t}  =&&  -6(E + PW) \frac{\partial \log\phi}{\partial t}
- \frac{1}{\phi^6} \frac{\partial}{\partial x^j}(\phi^6EV^j)\nonumber \\
&& -  P\biggl[\frac{\partial W}{\partial t} +
\frac{1}{\phi^6} \frac{\partial}{\partial x^j}(\phi^6 W V^j)\biggr]~~.
\end{eqnarray}
Momentum conservation takes the form,
\begin{eqnarray}
\frac{\partial S_i}{\partial t}& = & -6 S_i \frac{\partial \log\phi}{\partial t}
- \frac{1}{\phi^6} \frac{\partial}{\partial x^j}(\phi^6S_iV^j)
-\alpha \frac{\partial P}{\partial x^i} \nonumber \\
& + & 2\alpha (D+ E + PW)( W - \frac{1}{W}) \frac{\partial \log\phi}{\partial x^i}
+ S_j \frac{\partial \beta^j}{\partial x^i} \nonumber \\
& - & W (D + E + PW) \frac{\partial \alpha}{\partial x^i}  - \alpha W
(D+\Gamma E) {\partial \chi \over \partial x^i}~~. 
\label{hydromom}
\end{eqnarray}
where the last term in Eq. (\ref{hydromom}) is the  contribution from
the radiation reaction potential  $\chi$ as defined in \cite{wmm96,Lan16}. In the construction of quasi-stable orbit initial conditions,  this term is set to zero.
Including this term would allow for a calculation of the orbital evolution via gravity-wave emission in the CFA.  However, there is no guarantee that this 
is a sufficiently accurate solution  to the exact Einstein equations.  Hence, the CFA is most useful for the construction  of initial conditions.

 \subsection{Angular momentum and orbital frequency}
In the quasi-circular orbit approximation (neglecting angular momentum in the radiation field), this system has a Killing vector corresponding to rotation in the orbital plane.  Hence, for these calculations 
the angular momentum is well defined and given by an integral over the space-time components
of the stress-energy tensor \cite{mtw}, i.e.,
\begin{equation}
J^{i j} = \int ( T^{i 0} x^j - T^{j 0} x^i ) dV ~~.
\end{equation}
Aligning the $z$ axis with the angular momentum vector then gives,
\begin{equation}
J  = \int ( x S^y - y S^x ) dV ~~.
\end{equation}

 To find the orbital frequency detected by a distant observer corresponding to a fixed angular momentum we  employ a slightly modified derivation
of the orbital frequency than that of \cite{wmbook}.
In asymptotically flat coordinates the angular frequency detected by  a distant observer is simply
the coordinate angular velocity, i.e., one evaluates
\begin{equation}
\omega \equiv \frac{d \phi}{dt} = \frac{U^\phi}{U^0} ,
\end{equation}

In the ADM conformally flat (3+1) curved space, our only task is then to deduce $U^{\phi}$
from code coordinates.
For this we make a simple polar coordinate transformation keeping our conformally flat coordinates,
so
\begin{equation}
U^{\phi} = \Lambda^{\phi}_{\nu} U^{\nu} = \frac{x U^y - y U^x}{x^2 + y^2}
\end{equation}
Now, the code uses covariant four velocities,
$U_i = g_{i \nu} U^{\nu} = \beta_i U^0 + \phi^4 U^i$.
This gives $U^i = U_i \beta_i (W/\alpha) / \phi^4$.
Finally, one must density weight and volume average $\omega$ over the fluid  differential volume elements.
This gives,
\begin{equation}
\omega = \frac{\int d^3 x \phi^2 (D + \Gamma E) [(\alpha/W) (x U_y - y U_x) - (x \beta_y - y \beta_x) ]/(x^2 + y^2)}
        {\int d^3 x \phi^6 (D + \Gamma E)} .
\end{equation}
This form differs slightly from that of \cite{wmbook}, but leads to very similar results.

A key additional ingredient, however, is the implementation of a grid three velocity $V^{i}_{G}$
that minimizes the matter motion with respect to $U_i$ and $\beta_i$.
Hence, the total angular frequency to a distant observer
$\omega_{tot} = \omega + \omega_G$, and in the limit of rigid co-rotation,
$\omega_{tot} \rightarrow \omega_G$, where $\omega_G = x V^y + y V^x$.

For the orbit calculations illustrated  here we model corotating stars, i.e. no
spin in the corotating frame.  This minimizes matter motion on the grid.  However, we note that there is need at the present time of initial conditions for 
arbitrarily spinning neutron stars and the method described here is equally capable of supplying those initial conditions.

As a practical approach the simulation \cite{wmm96} of initial conditions is best run  first  with viscous damping in the hydrodynamics
for sufficiently long time (a few thousand cycles) to relax  the stars to a steady state.
One can  then run with no damping. In the present illustration we  examine stars at large separation that  are
in quasi-equilibrium circular orbits and stable hydrodynamic configurations.
These orbits span the time from  the last several minutes up to orbit inspiral.
Here, we illustrate  the stability of the multiple orbit hydrodynamic simulation and examine where the initial conditions for the strong field orbit dynamics computed here
deviates from the post-Newtonian regime.

\section{Code validation}

\subsection{Code Tests}\label{codetest}

\begin{figure}
\includegraphics[scale=0.5,angle=270]{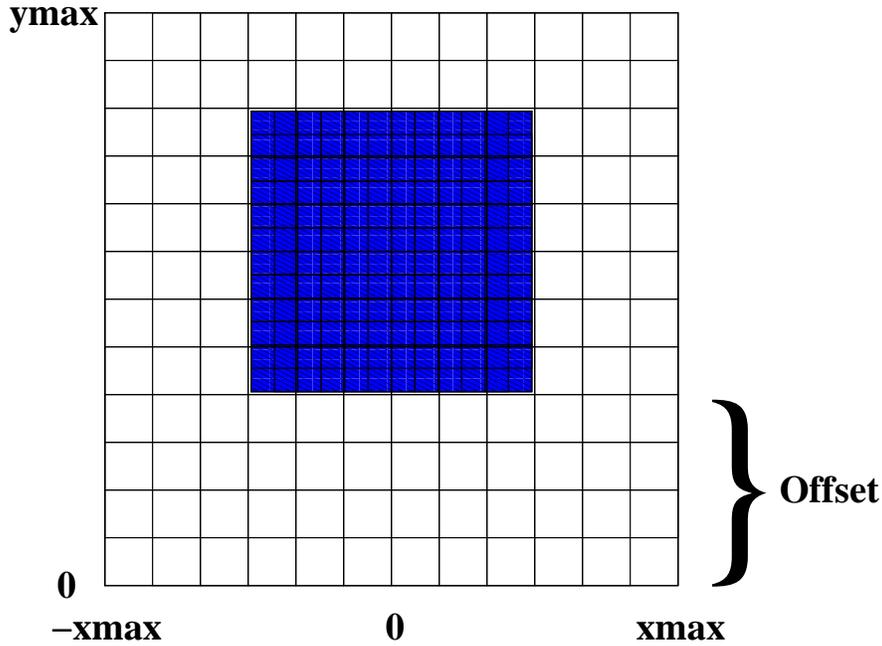}
\caption{Schematic representation of the field and hydrodynamics grid used in
the simulation. The inner blue grid represents the higher resolution matter grid
and the outer white grid represents the field grid.  The offset will be small
for small separations and large for large separations.\label{simgrid}}
\end{figure}

To evolve  stars at large separation distance it is best \cite{wmbook} to decompose  the grid into
a high resolution domain with a fine matter grid around the stars and a coarser
domain with an extended grid for the fields.  Figure~\ref{simgrid} shows a
schematic of this decomposition from \cite{haywood}.  

As noted in \cite{wmbook} it is best to keep the number of zones across each star between $25$
and $40$ \cite{haywood}.  This keeps the error in the numerics below $0.5\%$.  

\begin{figure}
\begin{center}
\centerline{\includegraphics[scale=0.4,angle=270]{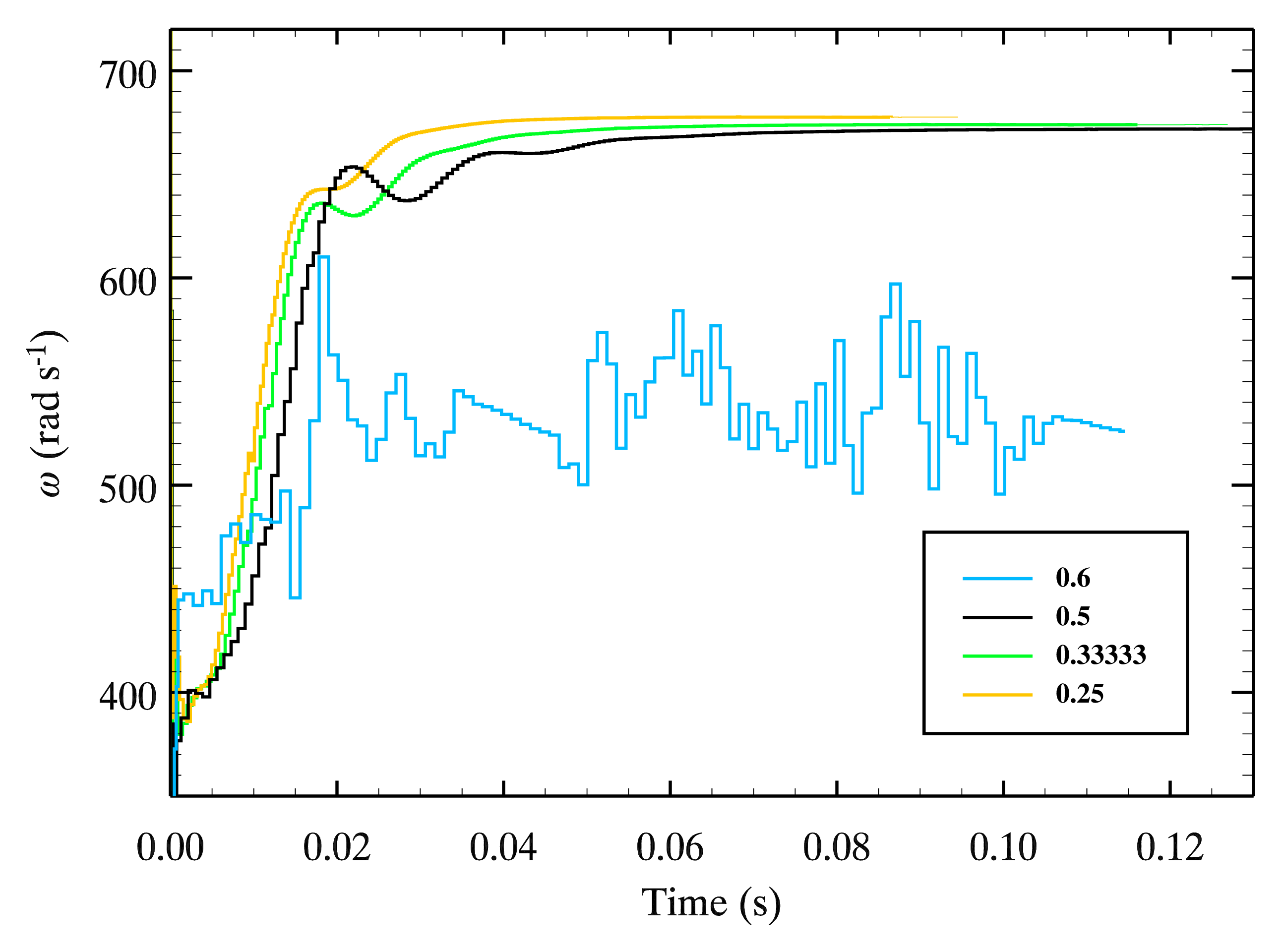}}
\caption{Comparison of the orbital angular velocity $\omega$ vs.~time for different values of the
Courant parameter $k$.  As can be seen, the simulations with $k=0.25-0.5$ result in
stable runs that converge to the same value, implying that a smaller $k$, or
equivalently a smaller $\delta t$, is not necessary and would only use extra CPU
time.  For comparison, we plot a simulation with $k=0.6$ to show that the
stability is lost for $k > 0.5$.\label{courcon}}
\end{center}
\end{figure}

\begin{figure}
\begin{center}
\centerline{\includegraphics[scale=0.4,angle=270]{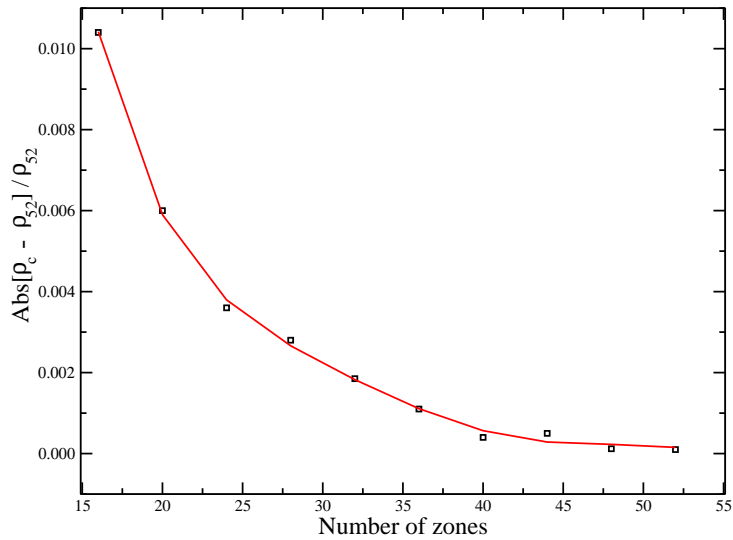}}
\caption{Plot of the error in the central density versus the number of zones
across the star.  It is clear that there is only a $1\%$ error with $\approx 15$
zones across the star.  Increasing the number of zones across the star so that
there are $> 35$ zones across the star produces less than a $0.1\%$
error.\label{denvszon}}
\end{center}
\end{figure}

It has also been pointed out \cite{wmbook} that  an artificial viscosity (AV) shock capturing scheme has an advantage 
over Riemann solvers in that only about half as many zones are required to accurately
resolve the stars when an AV scheme is employed compared to a Riemann solver.
Figure~\ref{courcon} shows a plot of orbital velocity vs.~time for various Courant parameters.  This figure establishes that the routines for the hydrodynamics
are stable (e.g. changing the Courant condition has little to effect) as long as $k < 0.5$. Figure~\ref{denvszon} illustrates the central density vs.~number of zones across the star when calculated with the MW EoS, i.e  the zero temperature, zero neutrino chemical potential EoS used to
model core-collapse  supernovae  \cite{wmm96,wmbook,maywil93}.

  This figure illustrates that here is only a $1\%$ error in central density with $\approx 15$
zones across the star, while increasing the number of zones across the star to $> 35$ produces less than a $0.1\%$.  In the illustrations below we maintain $k = 0.5$ and $\approx 25$ zones across each star as the best choice for both speed and accuracy needed to compute   stable orbital initial conditions.


\subsection{Orbit stability}
\begin{figure}
\begin{center}
\centerline{\includegraphics[scale=0.5,angle=0]{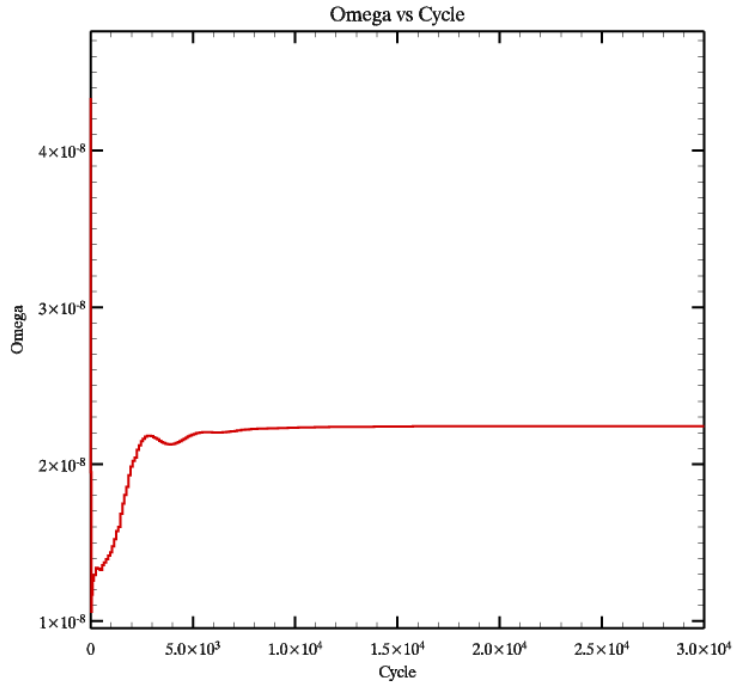}}
\caption{Plot of the orbital angular velocity, $\omega$, versus cycle.  When
$\omega$ stops changing with time the simulation has reached a circular binary
orbit solution. 
This run, which goes over $30,000$ cycles, lasts for $\approx 20$ orbits.
The geometrized unit of $\omega$ in simulation here is used.\label{shortrun}}
\end{center}
\end{figure}

As an illustration of the orbit stability Figure~\ref{shortrun} shows results from a simulation \cite{haywood}  in which  the angular momentum was fixed at  $J=2.7\times10^{11}$ cm$^{2}$
and the Courant parameter set to  $k=0.5$.  For this orbital calculation  the MW EoS was employed and
each star was fixed at a baryon mass of $M_B = 1.54$ M$_\odot$ and a gravitational mass
in isolation of $M_G = 1.40$~M$_\odot$.  

Figure~\ref{shortrun} shows the evolution of the orbital
angular velocity $\omega$, versus computational cycle for the first 30,000 code cycles corresponding to $\approx 20$ orbits. 
The stars were initially placed on the grid using a solution of the TOV equation
in isotropic coordinates for an isolated star.
The stars were initially  set to be corotating but were  allowed to settle into their binary equilibrium.
Notice that is takes $\sim 5,000$ cycles, corresponding to $\sim 3$ orbits,
just to approach the quasi-equilibrium binary solution.  Indeed, the stars continued to gradually compact and slightly increase in orbital frequency until $\sim 10 $ orbits, afterward, the stars were completely stable.
This particular figure extends to  $\approx 20$ orbits.   

\begin{figure}
\begin{center}
\centerline{\includegraphics[scale=0.5,angle=0]{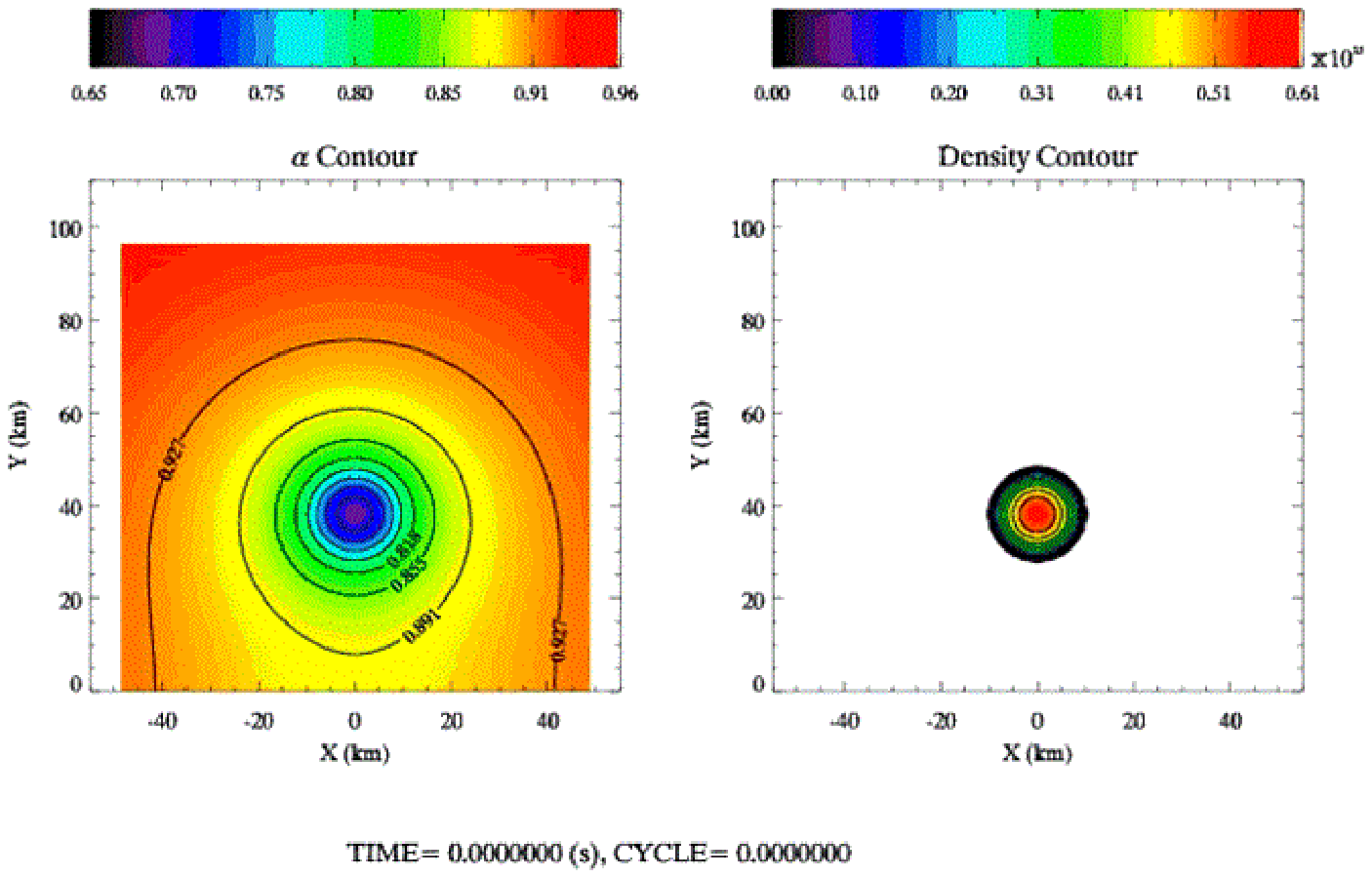}}
\centerline{\includegraphics[scale=0.5,angle=0]{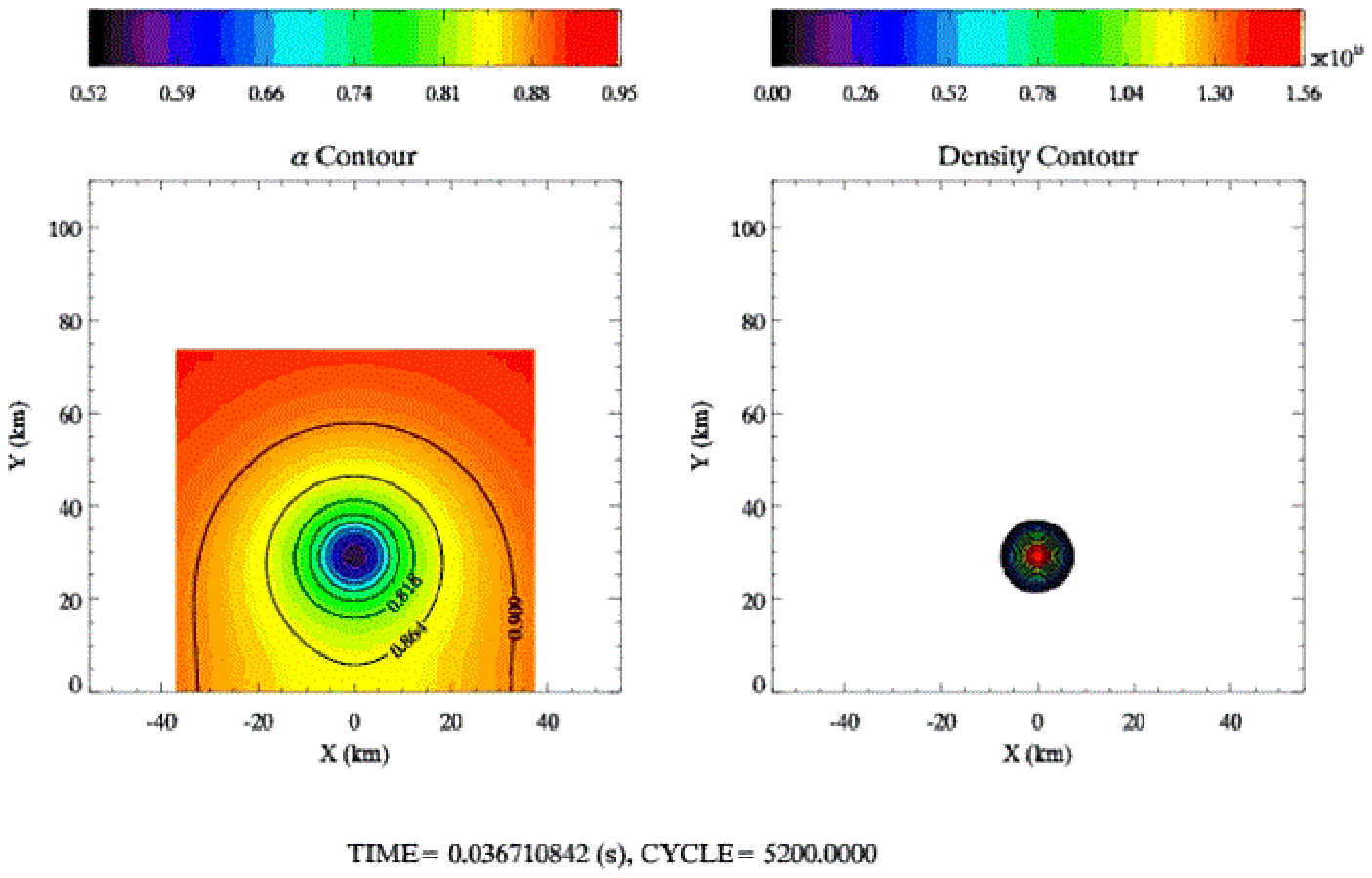}}
\centerline{\includegraphics[scale=0.5,angle=0]{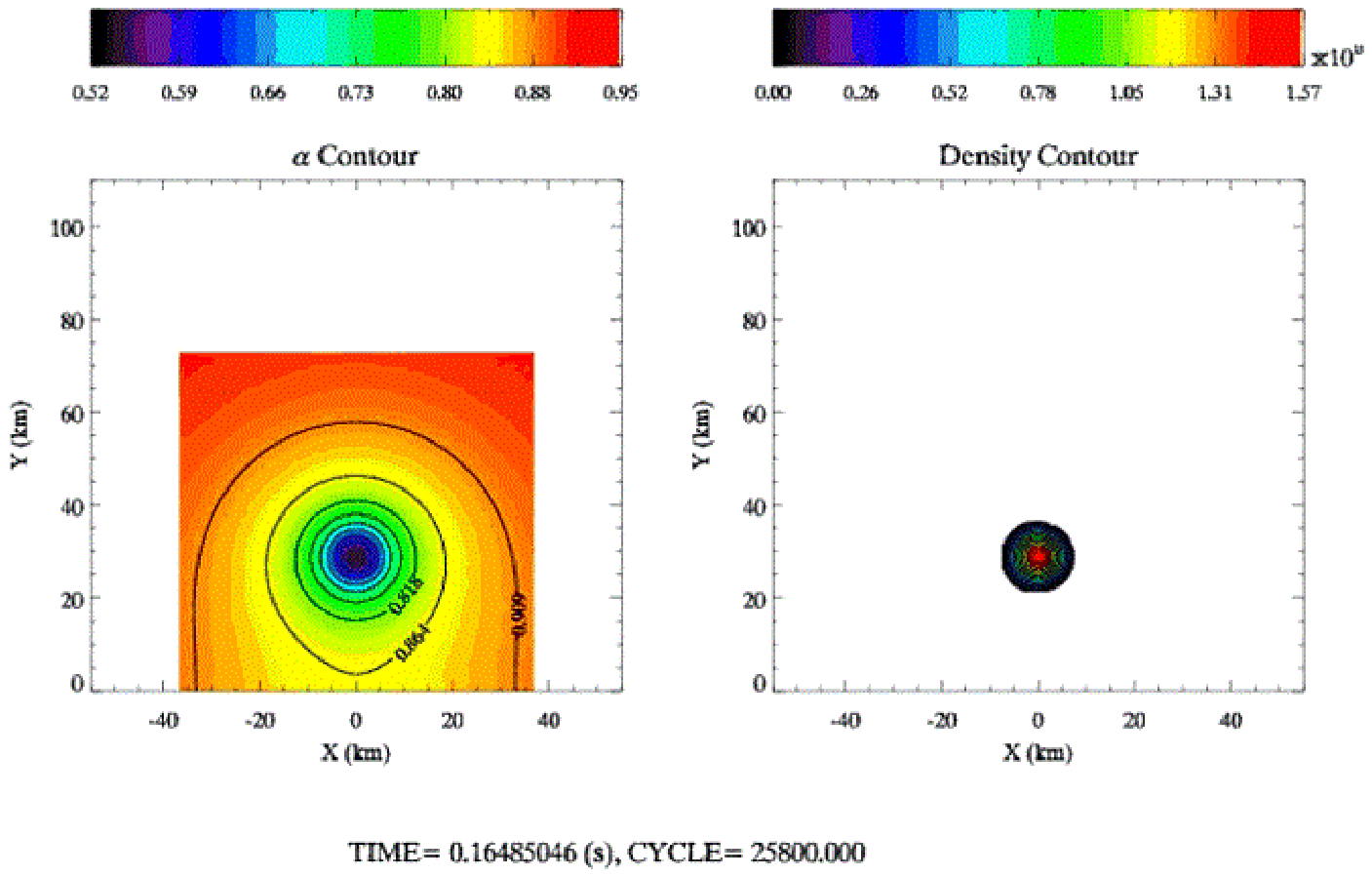}}
\caption{Contours of the lapse function (left) and central density (right) at cycle numbers 0 (top) , 5,200 (middle) , and 25,800 (bottom) corresponding to
roughly 0, 5, and 19 orbits. \label{alpha-den}}
\end{center}
\end{figure}

\begin{figure}
\begin{center}
\centerline{\includegraphics[scale=0.5,angle=0]{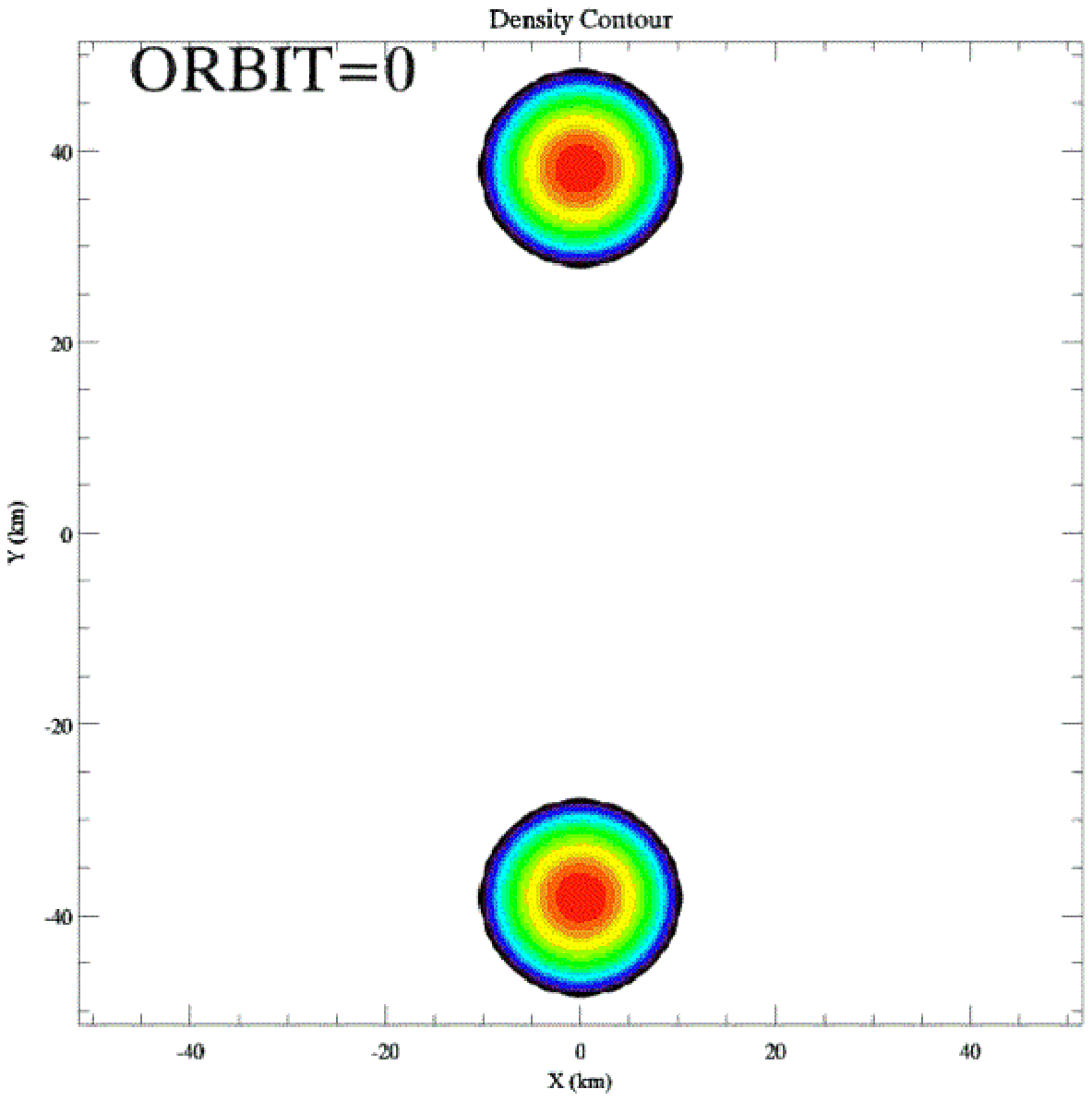}}
\centerline{\includegraphics[scale=0.5,angle=0]{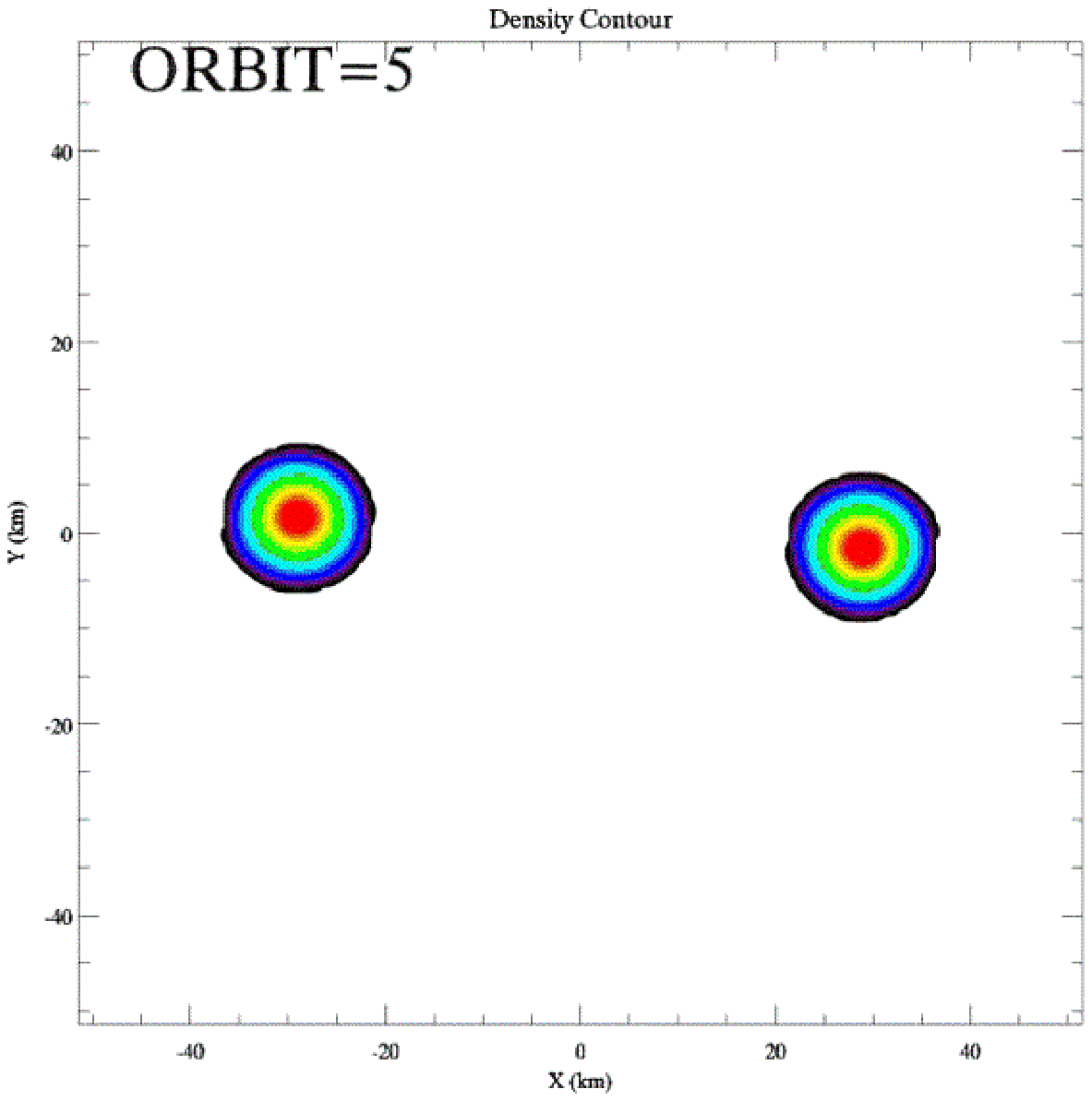}}
\centerline{\includegraphics[scale=0.5,angle=0]{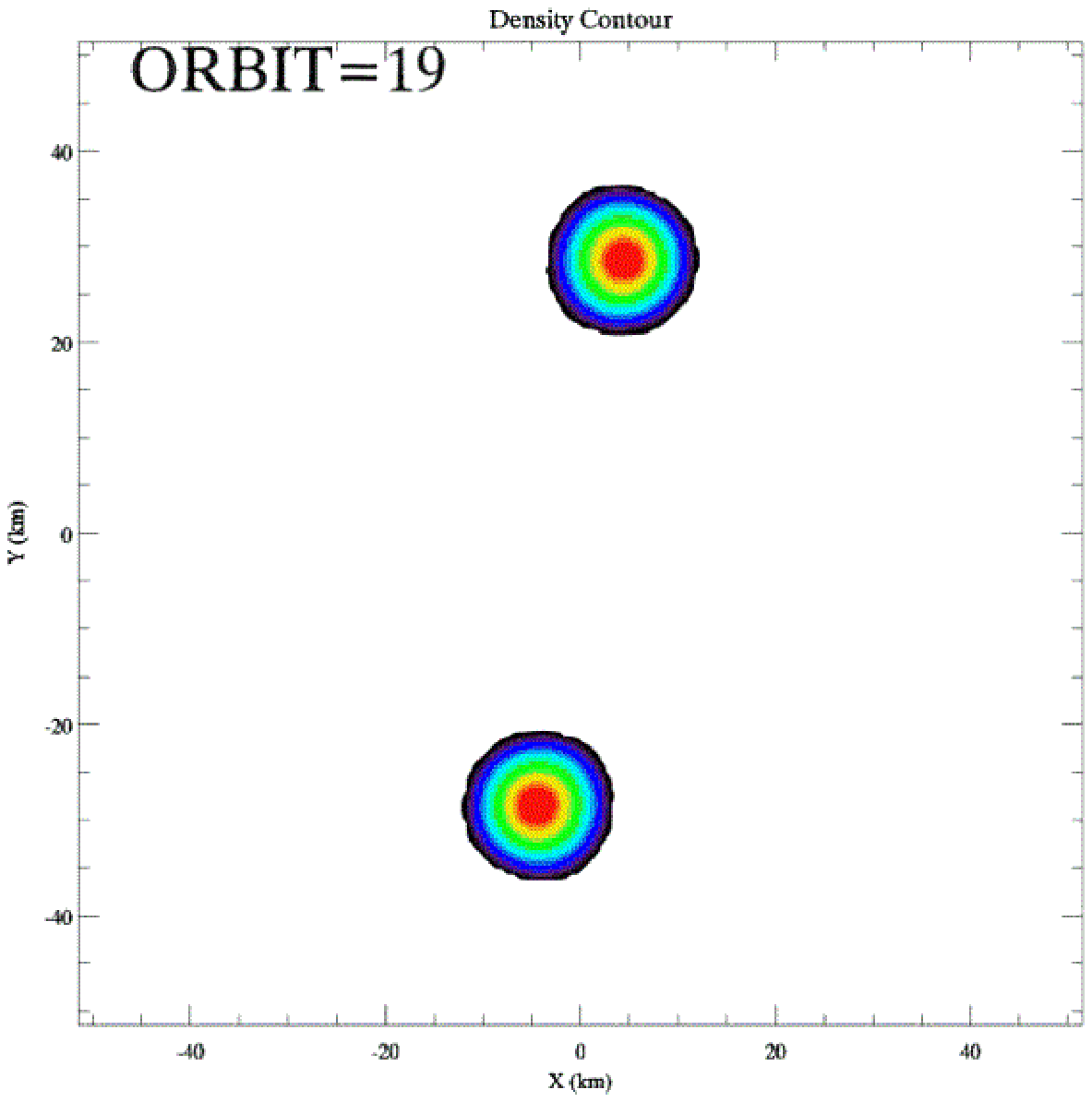}}
\caption{Contours of the central density for the binary system at the approximate number of orbits as labelled.\label{den-contours}}
\end{center}
\end{figure}

Fig. \ref{alpha-den} shows the contours of the lapse function $\alpha$ (roughly corresponding to the gravitational potential) and corresponding density profiles at  cycle numbers, 0, 5200, and 25800 ($\approx$ 0, 5, and 19 orbits). Figure~\ref{den-contours} shows the contours of
central density  and the orientation of the binary orbit corresponding to these cycle numbers. One can visibly see from these figures the relaxation of the stars after the first few orbits, and the stability of the density profiles after multiple orbits.
%
%

%
%

\begin{figure}
\begin{center}
\centerline{\includegraphics[scale=0.5,angle=0]{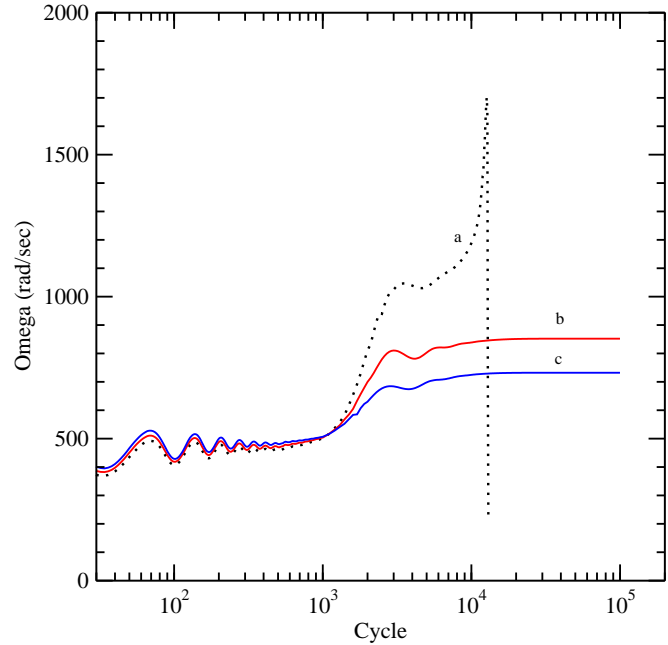}}
\caption{Plot of the orbital angular velocity, $\omega$, versus cycle.  When
$\omega$ stops changing with time the simulation has reached a circular binary
orbit solution. The run ($a$ $J = 2.7\times 10^{11}$ cm$^2$) goes over $\sim 10$ obits and then becomes  unstable to inspiral and merger after $\sim 10^4$ cycles.  
The stable two runs ($b$ for $J = 2.8\times 10^{11}$ cm$^{-2}$ and $c$ for $J = 2.9 \times 10^{11}$ cm$^{-2}$), were run for  $100,000$ cycles, and  $\approx 100$ orbits.
\label{longtrun}}
\end{center}
\end{figure}

\begin{figure}
\begin{center}
\centerline{\includegraphics[scale=0.5,angle=0]{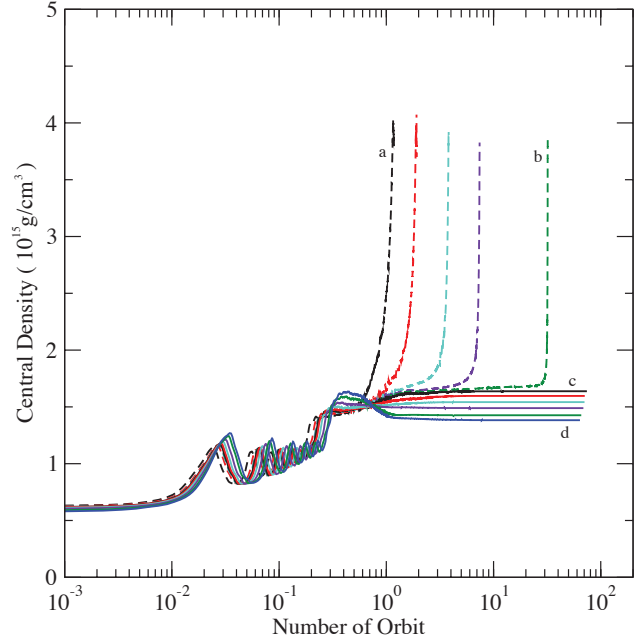}}
\caption{Plot of the central density, $\rho_c$, versus the number of orbit.
The dashed lines from left ($a$) to right ($b$) correspond to $J = 2.0, 2.2, 2.4, 2.6, 2.8 \times 10^{11}$ cm$^2$ and 
the solid lines from top ($c$) to bottom ($d$) are for $J = 3.0, 3.2, 3.4, 3.6, 3.8, 4.0 \times 10^{11}$ cm$^2$.  
The case of $J = 2.8 \times 10^{11}$ cm$^2$ shows stable obits until $\sim 30$.
\label{theta-rhoc}}  

\end{center}
\end{figure}

We note, however, that this orbit is  on the edge of the ISCO.  As such it could be  unstable to inspiral even after many orbits. Figures \ref{longtrun} and \ref{theta-rhoc} further  illustrate this point.  In these simulations various angular momenta  were computed with a slightly higher neutron-star mass ($M_b = 1.61$ M$_\odot$, $M_g = 1.44$ M$_\odot$), but the same MW EoS.  In this case the orbits  were followed for nearly 100 orbits.    

Figure \ref{longtrun} illustrates orbital angular frequency vs.~cycle number for three representative angular momenta bracketing the ISCO.  
The orbital separation for the lowest angular momentum ($J = 2.7 \times 10^{11}$ cm$^{-2}$) shown on Figure \ref{longtrun} is just inside the ISCO.  Hence, even though it requires about 10 orbits before inspiral, the orbit is eventually unstable. 

Similarly, Figure \ref{theta-rhoc} shows the central density vs.~number of orbits for 11 different angular momenta, five of which have orbits inside the ISCO.
Here one can see that only orbits with $J \ge 3.0 \times 10^{11}$ cm$^{-2}$ are stable.  Indeed, for these cases, after about the first 3 orbits the orbits continue with  almost no discernible change in orbit frequency or central density.  


As mentioned previously, the numerical relativistic neutron binary simulations of
\cite{duez03, Marronetti03, Miller04, Miller05, Miller07, Uryu06, Kiuchi09, Bernuzzi15, DePietri15} all start with initial data that are
subsequently evolved in a different manner than those with which they were created.
One conclusion that may be drawn from the above set of simulations, however,  is that the initial data  
must be evolved for ample time ($>3$ orbit)  for the stars to reach a true quasi-equilibrium
binary configuration.  That has not always been done in the literature.

\section{Sensitivity of initial condition orbital parameters to the equation of state}

\subsection{Equations of State}
One hope in the forthcoming detections of gravitational waves is  that a sensitivity exists to  the neutron star equation of state.   
For illustration in this review we utilize several representative  equations of state often employed in the literature.  These span a range from relatively soft to  stiff nuclear matter.
These are used to illustrate the EoS dependence of the initial conditions.
One EoS often employed  is that of a polytrope,
i.e., $p=K \rho^{\Gamma}$, with $\Gamma=2$, where
in cgs units,  $K=0.0445(c^{2}/\rho_{n})$, and
$\rho_{n}=2.3\times 10^{14}$ g cm$^{-3}$.  These parameters,
with $\rho_{c}=4.74\times 10^{14}$ g cm$^{-3}$, produce an isolated star having
radius $=17.12$ km and baryon mass $= 1.5$  M$_{\odot}$.
As noted in previous sections  we utilize  the zero temperature, zero neutrino chemical potential MW EoS \cite{wmm96,wmbook,maywil93}.
The third is the equation of state developed by Lattimer and Swesty \cite{ls91}
with two different choices of compressibility, one having compressibility $K=220$ MeV, and
the other having $K=375$ MeV.
We denote these  as LS 220 and LS 375.
The fourth EoS has been developed by Glendenning \cite{glen96}. This EoS has $K=240$ MeV,
which is close the experimental value  \cite{garg04}. We denote this EoS as GLN.
Table~\ref{isostars} illustrates \cite{haywood} the properties of isolated neutron stars generated with
each EoS. For each case  the baryon mass was chosen to obtain a gravitational mass for each star
of $1.4 M_{\odot}$.

\begin{figure}[tpb]
\includegraphics[scale=0.5,angle=270]{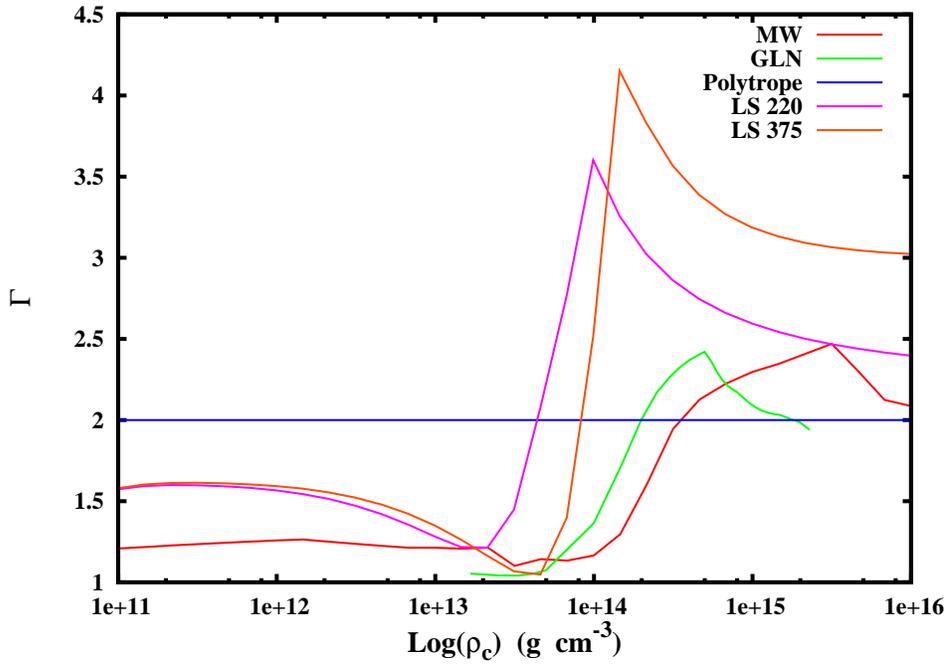}
\caption{EoS index $\Gamma$ vs.~central density for various equations of state.
Large $\Gamma$ implies a stiff EoS.}
\label{gvden}
\end{figure}

\begin{table}
\caption{\uppercase{Table presenting central density, baryon mass, and
gravitational mass for the five adopted equations of state}\label{isostars}}
\begin{center}
\begin{tabular}{cccc}
\hline
\hline
EoS   &$\rho_c$  ($\times 10^{15} g~cm^{-3}$)                  &$M_{B}$  $(M_{\odot}$)     &$M_{G}$ $(M_{\odot}$) \\
\hline
$\Gamma=2$ Polytrope &   0.474              & 1.50         & 1.40\\
MW    &1.39                       &1.54        &1.40 \\
LS 220&0.698                      &1.54        &$~~ \sim 1.40$\\
LS 375&0.492                      &1.54        &$~~ \sim 1.40$\\
GLN   &1.56                       &1.54        &1.40 \\
\hline
\end{tabular}
\end{center}
\end{table}
In Fig.~\ref{gvden} we plot the equation of state index, $\Gamma$ versus density,
$\rho$, for the various EoS's considered here. These are compared the simple polytropic $\Gamma$ = 2 EoS
often employed in the literature.

 \subsection{EoS dependence of the Initial Condition Orbit Parameters}
 Table~\ref{jvals} summarizes the initial condition orbit parameters \cite{haywood} at various fixed angular momenta for the various equations of state.  In the case of orbits unstable to merger, this table lists the orbit parameters just before inspiral.  
These orbits span a range in specific angular momenta $J/M_0^2$ of $\sim 5$ to 10.  
The equations of state listed in Table \ref{jvals} are in approximate order of increasing stiffness from the top to the bottom.  

As expected, the central densities are much higher for the relatively soft MW and GLN equations of state.  Also, the orbit angular frequencies are considerably lower for the extended mass distributions of the stiff equations of state than for the more compact soft equations of state.  These extended mass distributions induce a sensitivity  of the emergent gravitational wave frequencies and amplitude due to the strong dependence of the gravitational wave frequency to the quadrupole moment of the mass distribution.

\begin{table}
\caption{Orbital parameters for each EoS\label{jvals}}
\begin{tabular}{ccccccccc}
\hline
\hline
{EoS} &  {$J (cm^{2})$} &  {$\omega (rad~s^{-1})$} &
 {$d_{p} (km)$} &  {$d_c (km)$} &
 {$M_{ADM} (M_{\odot})$} &  {$\rho_{c} (g~cm^{-3})$} \\
\hline
MW&$2.6\times 10^{11}$&780.92&65.22&51.52&1.391&$1.67\times10^{15}$\\
&$2.7\times 10^{11}$&671.85&71.18&57.24&1.393&$1.62\times10^{15}$\\
&$2.8\times 10^{11}$&602.80&76.94&61.86&1.394&$1.60\times10^{15}$\\
&$3.0\times 10^{11}$&482.30&86.91&72.36&1.396&$1.55\times10^{15}$\\
&$3.5\times 10^{11}$&300.46&116.13&100.8&1.399&$1.44\times10^{15}$\\
&$3.8\times 10^{11}$& 235.72 & 136.93& 119.74 & 1.401& $1.39\times10^{15}$ \\
\hline
GLN&$2.7\times 10^{11}$&666.5&71.62&57.67&1.390&$1.73\times10^{15}$\\
&$2.8\times 10^{11}$&592.34&77.82&62.81&1.391&$1.69\times10^{15}$\\
&$3.0\times 10^{11}$&475.05&88.06&73.53&1.394&$1.61\times10^{15}$\\
&$3.2\times 10^{11}$&391.75&100.34&84.31&1.396&$1.56\times10^{15}$\\
\hline
LS 220&$2.7\times 10^{11}$&523.59&90.77&77.34&1.403&$7.18\times10^{14}$\\
&$2.8\times 10^{11}$&472.08&97.53&83.08&1.404&$7.14\times10^{14}$\\
&$3.0\times 10^{11}$&389.96&109.78&94.84&1.405&$7.06\times10^{14}$\\
&$3.2\times 10^{11}$&327.04&122.51&107.10&1.407&$6.98\times10^{14}$\\
\hline
LS 375&$2.7\times 10^{11}$&490.09&97.09&83.92&1.404&$5.00\times10^{14}$\\
&$2.8\times 10^{11}$&442.40&103.95&90.04&1.405&$4.98\times10^{14}$\\
&$3.0\times 10^{11}$&366.67&116.65&102.50&1.406&$4.95\times10^{14}$\\
&$3.2\times 10^{11}$&307.80&130.72&115.60&1.407&$4.92\times10^{14}$\\
\hline
Polytrope&$1.8\times 10^{11}$&804.70&63.30&51.20&1.395&$6.78\times10^{14}$\\
&$2.1\times 10^{11}$&826.03&67.85&55.18&1.396&$7.00\times10^{14}$\\
&$2.3\times 10^{11}$&762.37&74.64&61.72&1.397&$6.55\times10^{14}$\\
&$2.5\times 10^{11}$&624.33&85.87&72.71&1.399&$6.24\times10^{14}$\\
&$2.6\times 10^{11}$&532.83&94.04&80.45&1.400&$6.17\times10^{14}$\\
&$2.7\times 10^{11}$&477.19&101.34&86.95&1.400&$6.05\times10^{14}$\\
\hline
\end{tabular}
\end{table}

\subsection{ Gravitational Wave Frequency}
The physical processes occurring during the last orbits of a neutron star binary
are currently a subject of intense interest. As the stars approach their final
orbits it is expected that the coupling of the orbital motion to the hydrodynamic evolution
of the stars in the strong relativistic fields could provide insight into
various  physical properties of the coalescing system \cite{mw97,cut93}.
In this regard, careful modeling of the initial conditions is needed which includes both the nonlinear general relativistic and
hydrodynamic effects as well as a realistic neutron star equation of state.

\begin{figure}
\includegraphics[scale=0.5,angle=270]{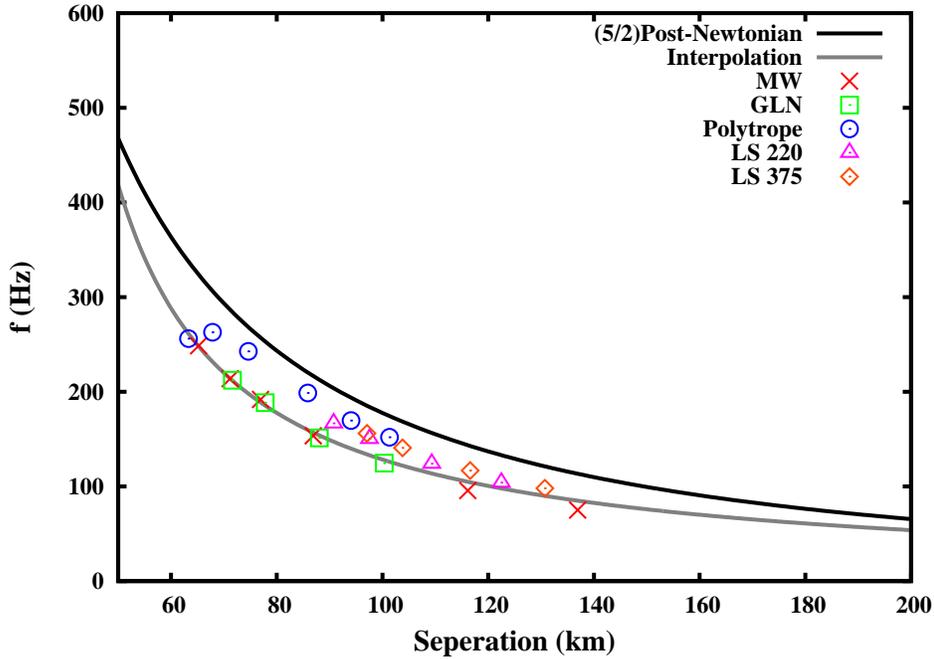}
\caption{Computed gravitational wave frequency, $f$, versus proper separation
for each EoS as labelled.  The black line corresponds to the (post)$^{5/2}$-Newtonian
estimate.  Frequencies obtained from the stiff and polytropic equations of
state do not deviate by more than $\sim 10$\% from the PN prediction until a frequency greater than $\sim 300$ Hz.  The gray line is an extrapolation
of the frequencies obtained using the soft MW and GLN EoSs.  These begin to deviate by more than 10\%  from the PN prediction at a frequency of $\sim 100$ Hz.\label{fvd}}
\end{figure}

Fig.~\ref{fvd} shows the EoS sensitivity of the gravitational wave frequency   $f = \omega/\pi$  as a function of proper
separation $d_p$ between the stars for the various orbits and equations of state summarized in Table \ref{jvals}.
 These are compared with the circular orbit
condition in the (post)$^{5/2}$-Newtonian, hereafter PN, analysis of reference
\cite{kidder}. In that paper a search was made for the inner most stable
circular orbit in the absence of radiation reaction terms in the equations of
motion.  This is analogous to the calculations performed here which also analyzes
orbit stability  in the absence of radiation reaction.

In the (post)$^{5/2}$-Newtonian equations of motion, a circular orbit is derived
by setting time derivatives of the separation, angular frequency, and the radial
acceleration to zero. This leads to the circular orbit condition \cite{kidder},
\begin{equation}
\omega_0^2 = m A_0/d_h^3~~,
\label{circular}
\end{equation}
where $\omega_0$ is the circular orbit frequency and  $m = 2M_G^0$, $d_h$ is the
separation in harmonic coordinates, and $A_0$ is a relative acceleration
parameter which for equal mass stars becomes,
\begin{equation}
A_0 = 1 - \frac{3}{2} \frac{m}{d_h}\biggl[3 - \frac{77}{8} \frac{m}{d_h} +
(\omega_0 d_h)^2 \biggr] + \frac{7}{4}(\omega_0  d_h)^2~~.
\label{a0}
\end{equation}
Equations (\ref{circular}) and (\ref{a0}) can be solved to find the orbit
angular frequency as a function of harmonic separation $d_h$. The gravitational wave
frequency is then twice the orbit frequency,
$f = \omega_0/\pi$.

Although this is a  gauge dependent comparison, for illustration we show  in Fig.~\ref{fvd}  the  calculated
gravitational wave frequencyis compared to the PN expectation  as a function of proper  binary separation distance up to $ 200$
km.  One should keep in mind, however, that there is some uncertainty in associating our proper distance with the
PN parameter ($m/r$).  Hence , a comparison with the PN results is meaningless.  It is nevertheless instructive to consider the difference in the numerical results as the stiffness  of the EoS is varied.  THe polytropic and stiff EoS's  begin to deviate (by $> 10$\%)  from the softer equations of state (MW and GLN)  for a gravitational wave frequency as low as $\sim 100$ Hz and more or less continue to deviate as the stars approach the ISCO at higher frequencies.

Indeed,  a striking feature of Figure \ref{fvd} is that as the stars approach the ISCO, the frequency
varies more slowly with diminishing separation distance  for the softer equations of state.  A  gradual change in frequency can mean more orbits in the LIGO window, and hence, 
a stronger signal to noise (cf. \cite{Lan16}).

Also, for a soft EoS the orbit becomes unstable to
inspiral at a larger separation.   At least part of the difference between the soft and stiff EoSs can be
attributed to the effects of the finite size of the stars which is more compact for the soft equations of state.  That is the stars in a soft EoS are more point-like \cite{Read13}.

We note that, for comparable angular momenta, our results are consistent with
the EoS sensitivity study of \cite{Read13} based upon 
a set of equations of state parameterized by a segmented polytropic indices and an overall pressure scale.  
Their calculations, however, were based upon two independent numerical relativity codes.  
The similarity of their simulations  to the  results in Table \ref{jvals} further confirms the broad validity of the CFC approach when applied to initial conditions.
For example, their orbit parameters are summarized in Table II of \cite{Read13}.  Their softest EoS is the Bss221 which corresponds to an adiabatic index of $\Gamma = 2.4$ for the core, and a baryon mass of 1.501 M$_\odot$ and an ADM mass of 1.338 M$_\odot$ per star for a specific angular momentum of $1.61 \times 10^{11}$ cm$^{2}$ (in our units) with a corresponding gravitational wave frequency of 530 Hz at a proper separation of 46 km.    This EoS is comparable to the polytropic, MW and GLN EoSs shown on Fig.~\ref{fvd}.  For example, our closest orbit with the  $\Gamma = 2$ polytropic EoS corresponds to a specific angular momentum of $1.8 \times 10^{11}$ cm$^2$ and an ADM mass of 1.39 M$_\odot$compared to their ADM mass of 1.34 M$_\odot$ at $J = 1.6 \times 10^{11}$ cm$^2$ for the same  baryon mass of 1.5 M$_\odot$.  Although for the softer EoSs, their results are  for a closer orbit than the numerical points given on Figure \ref{fvd}, an extension of the grey line fit to the numerical simulations of the soft EoSs would predict a frequency of 540 Hz at the same proper separation of 46 km compared to 530 Hz in the Bss221 simulation of \cite{Read13}.

The main parameter characterizing the last stable orbit in the post-Newtonian
calculation is the ratio of coordinate separation to total mass (in isolation)
$d_h/m$. The analogous quantity in our non-perturbative simulation is proper
separation to gravitational mass, $d_P/m$. The separation corresponding to the
last stable orbit in the post-Newtonian analysis does not occur until the stars
have approached 6.03 $m$. For $M_G^0 = 1.4 M_\odot$ stars, this would correspond to a
separation distance of about 25 km.  In the results reported here the last
stable orbit occurs somewhere just below 7.7 $m_G^0$ at a proper separation
distance of $d_P \approx 30$ km for both the polytropic and the MW stars.

\subsection{Binding Energy}

A somewhat less gauge dependent  quantity that may be compared with PN initial conditions solution is the binding energy.
 The binding energy of an isolated star is defined as
\begin{equation}
\label{beni}
E_{b}=M_{g}-M_{0}.
\end{equation}
The total binding energy of the system is defined as
\begin{equation}
\label{bent}
E_{t}=M-2 M_{0}.
\end{equation}
In Equations~(\ref{beni}) and (\ref{bent}), $M_g$ is the ADM mass of a spherical star in
isolation and $M_{0}$ is the baryon mass.  Also of interest is $M_{t}=2 M_g$.
$M$ is the ADM mass of the binary system and will be different from $M_{t}$ due
to the binding energy between the  stars \cite{shibu01}.  The
(post)$^2$-Newtonian approximation to the binding energy is given by
\cite{shibu01},
\begin{equation}
\displaystyle E_{2PN}=-\eta M_{g} v^2\left (1-\frac{9+\eta}{12}v^2-\frac{81-57
\eta+\eta^2}{3}v^4 \right )+2 E_{b}.
\end{equation}
Here, $\eta$ is the ratio of the reduced mass to $M_{t}$ ($\eta = 1/4$ for equal mass
binaries) and $v=(M_{t} \omega)^{1/3}$, where $\omega$ is the orbital angular
velocity.

The (post)$^{3}$-Newtonian approximation has also been derived \cite{lb3pn,Blanchet14} and is
\begin{eqnarray}
 E_{3PN}=&-&\frac{mu v^{2}}{2} \biggl( 1+\left(-\frac{3}{4}-\frac{1}{12}\eta
\right )v^{2} \nonumber \\
&+&\left\{-\frac{27}{8}+\frac{19}{8}\eta-\frac{1}{24}\eta^{2}\right)v^{4}  \nonumber \\
&+& \biggl[-\frac{675}{64} +
 \left( \frac{209323}{4032}-\frac{205}{96}\pi^{2} - \frac{110}{9} \lambda \right) \eta \nonumber \\
&-& \frac{155}{96}\eta^{2}-\frac{35}{5184}\eta^{3 }\biggr] v^{6} \biggr)~~.
\end{eqnarray}

\begin{figure}
\begin{center}
\centerline{\includegraphics[scale=0.5,angle=270]{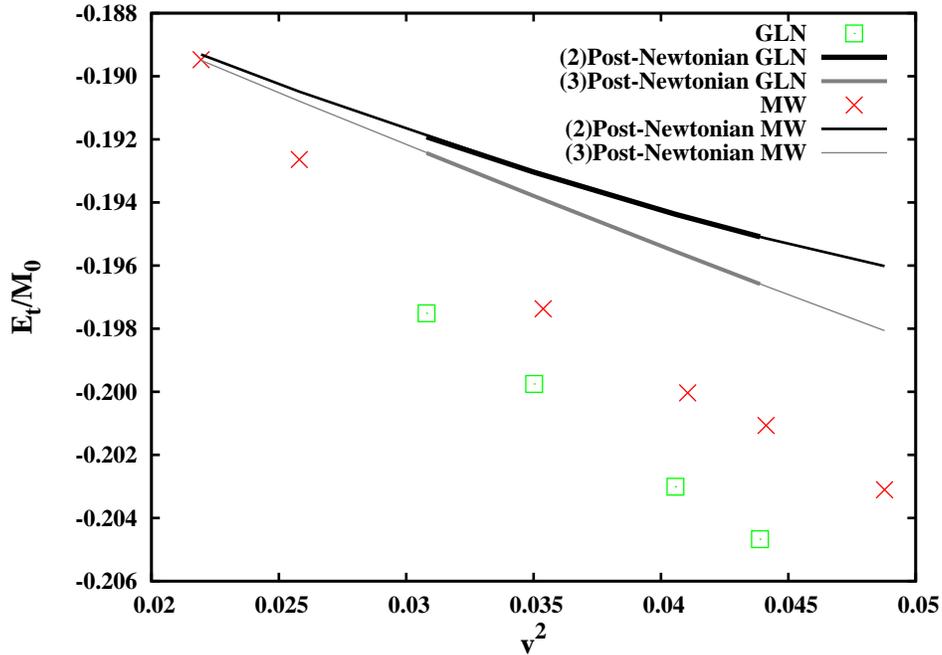}}
\caption{Plot of the total binding energy of the stars, $E_t$, versus the three-velocity $v^{2}$
for the MW and GLN EoS's.  This is compared with the 2nd and 3rd order PN prediction.  Notice that the 2nd- and 3rdPN approximation to the binding
energy is the same for both stars while the simulation begins to deviate for $v > 0.15$.\label{evmw}}
\end{center}
\end{figure}

\begin{figure}
\begin{center}
\centerline{\includegraphics[scale=0.5,angle=270]{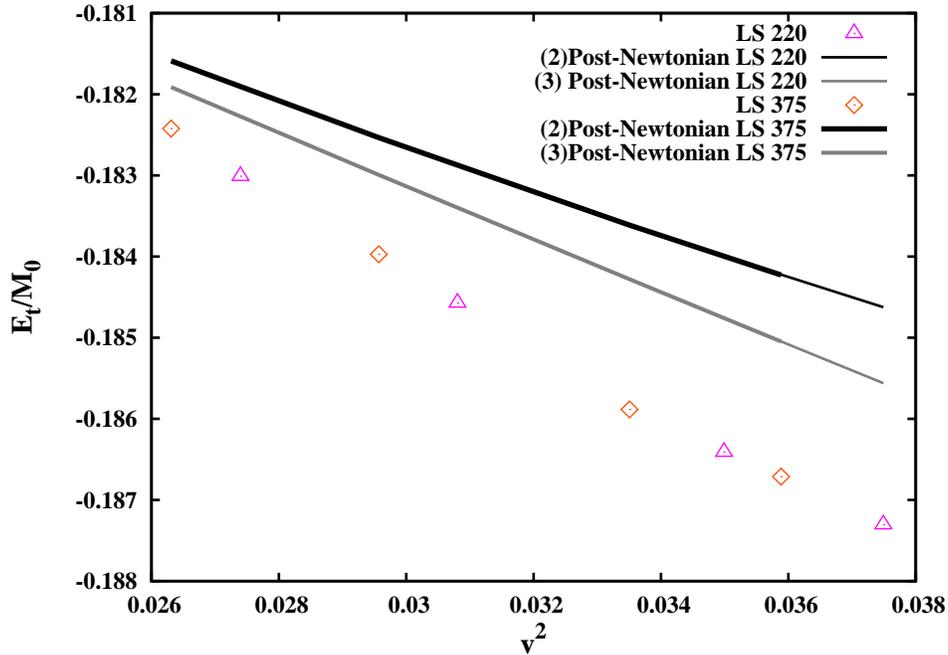}}
\caption{Same as Fig.~\ref{evmw} for the LS 220 and LS 375 EoS's. (Note the change of scale for the horizontal axis).\label{evls}}
\end{center}
\end{figure}

In Figures~\ref{evmw} and \ref{evls} we plot the total binding energy per baryon, 
$E_t/M_{0}$ versus $v^2$.  The simulations diverge from the PN results for  $v > 0.15$.  However, as expected,  extending from  from 2PN to
3PN diminishes the discrepancy. The simulations which use the MW EoS give the
same $E_t$ as the (post)$^{3}$-Newtonian solution at angular momentum
$J=3.8\times10^{11}$ cm$^{2}$.  Higher order corrections in the
PN expansion should bring agreement between the simulations and the PN expansion at
lower $J$ values.  This would require (post)$^{6}$-Newtonian order, where finite
size effects must be taken into account in the expansion \cite{lb3pn}.  Note
also, that even though the gravitational and baryon masses generated with the MW and
GLN EoS are the same (see
Fig.~\ref{fvd}), the resulting binary binding energies are different.    Since the gravitational wave frequencies are the same, but the binding
energies are different, it should be possible to distinguish the ``true'' EoS from
the gravitational wave signal which depends strongly on the mass distribution associated with a given binding energy.


\section{Conclusions}

The relativistic hydrodynamic equilibrium in the  CFA remains as a viable  approach to calculate the initial conditions for calculations of binary neutron stars.  
In this review we have illustrated that one must construct initial conditions that have run for at least several orbits before equilibrium is guaranteed.  We have demonstrated that beyond the first several orbits the equations are stable  over many orbits ($\sim 100$). 
We also have shown that such multiple orbit simulations are valuable as a means to estimate  the location of the ISCO prior to a full dynamical calculation.  
Moreover, we have  examined 
the sensitivity of the initial-condition orbit parameters and initial gravitational-wave frequency to the equation of state.   We have illustrated 
how the initial-condition orbital properties (e.g. central densities, orbital velocities, binding energies) and location of the ISCO are significantly effected by the stiffness of the EoS.

\section*{Acknowledgements}
Work at the University of Notre Dame (G.J.M.) supported by the U.S. Department of Energy
under Nuclear Theory Grant DE-FG02-95-ER40934, and by the University of Notre Dame Center for Research Computing.
One of the authors (N.Q.L.) was also supported in part by the National Science Foundation through
the Joint Institute for Nuclear Astrophysics (JINA) at UND, and in part by the Vietnam Ministry of Education (MOE).
N.Q.L. would also like to thank the Yukawa Institute for Theoretical Physics for their hospitality during a visit where part of this work was done.


\section*{References}

\begin{thebibliography}{99}
\bibitem{aligodetect} Abbot, P. B. et al. (LIGO Scientific Collaboration) 2016, Phys. Rev. Lett., 116, 061102

\bibitem{aLIGO} Asai, J. et al. (LIGO Scientific Collaboration) 2015 {\it Class. Quant. Grav.} {\bf  32} 074001 (2015).

\bibitem{VIRGO} Acernese  F et al. (VIRGO) 2015  {\it Class. Quant. Grav.} {\bf 32} 024001 

\bibitem{KAGRA} Somiya  K (KAGRA Collaboration) 2012  {\it Class. Quant. Grav.} {\bf  29} 124007 

\bibitem{ligoweb} LIGO Scientific Collaboration, http://www.ligo.org










\bibitem{thorne96} Thorne K S 1996 {\it Compact Stars in Binaries}, J. van Paradijs, E. P. J. van den Heuvel, and E. Kuulkers, editors, IAU Symp. {\bf 165} 153

\bibitem{Harry10} Harry G M 2010 {\it Class. Quant. Grav.} {\bf 27} 084006

\bibitem{shaprev} Rasio F A and Shapiro S L1999 {\it Class. Quant. Grav.} {\bf 16} 1

\bibitem{bail96} Bailes M 1996 {\it Compact Stars in Binaries}, J. van Paradijs, E. P. J. van den Heuvel, and E. Kuulkers, editors, IAU Symp. {\bf 165}  213

\bibitem{tutu93} Tutukov A V  and Yungelson, L R 1993 {\it MNRAS} {\bf 260} 675

\bibitem{phin91} Phinney  E S 1991 {\it Astrophys. J.} {\bf 380} L17

\bibitem{Kalogera04} Kalogera K et al., 2014 {\it Astrophys. J. Lett.} {\bf 614} L137 

\bibitem{Abadie10} Abadie J et al. {\it Class. Quant. Grav.} {\bf 27} 173001

\bibitem{burgay03} Burgay M et al.  2003  {\it Nature} {\bf 426} 531

\bibitem{Lattimer12} Lattimer J M 2012 {\it Ann. Rev. Nucl Part. Sci} {\bf 62} 485

\bibitem{apos96} Apostolatos T A  1996 {\it Phys. Rev. D} {\bf 54} 2421

\bibitem{droz97} Droz S and Poisson E 1997 {\it Phys. Rev. D} {\bf 56} 4449

\bibitem{sath00} Sathyaprakash B S 2000 {\it Class. Quant. Grav.} {\bf 17} L157

\bibitem{buo03} Buonanno A, Chen Y, and Vallisneri M 2003 {\it Phys. Rev. D} {\bf 67} 024016

\bibitem{bose05} Bose  S 2005  {\it Phys. Rev. D} {\bf 71} 082001

\bibitem{ott06} Ott C D,  Burrows A, Dessart L, and Livne E 2006 {\it Phys. Rev. Lett.} {\bf 96} 201102

\bibitem{Ajith14} Ajith P, Fotopoulos N, Privitera S, Neunzert A,  Mazumder N and Weinstein A J 2014, {\it Phys. Rev. D} {\bf 89} 084041

\bibitem{Pannarale15}  Pannarale F, Berti E,  Kyutoku K,  Lackey,B D and Shibata M  2015 {\it Phys. Rev. D} {\bf 92}081504

\bibitem{Agathos15}  Agathos M, Meidam J, Del Pozzo W, Li T G F, Tompitak M,  Veitch J, Vitale S, and Van Den Broeck C 2015
{\it Phys. Rev. D} {\bf 92} 023012  

\bibitem{Clark15} Clark, J A, Bauswein, A, Stergioulas, N and  Shoemaker D 2015, eprint arXiv:1509.08522

\bibitem{allen05} B. Allen B,  Anderson W G, Brady P R, Brown D A, and Creighton J D E 2012
        {\it Phys. Rev. D} {\bf 85} 122006

\bibitem{lb3pn} Blanchet L 2002 {\it Living Rev. Relativity} {\bf 5} 3

\bibitem{Blanchet14} Blanchet L 2014 {\it Living Rev. Relativity} {\bf 17} 2

\bibitem{Mishra15} Mishra C K, Arun K G and  Iyer B R 2015, {\it Phys. Rev. D} {\bf 91} 084040

\bibitem{wm95} Wilson J R and Mathews G J, 1995 {\it Phys. Rev. Lett.} {\bf 75} 4161

\bibitem{wmm96} Wilson J R, Mathews G J, and Marronetti P 1996 {\it Phys. Rev. D} {\bf 54} 1317

\bibitem{mmw98} Mathews G J, Marronetti P, and Wilson J R 1998  {\it Phys. Rev. D} {\bf 58} 043003

\bibitem{mw00} Mathews G J, and Wilson J R  2000 {\it Phys. Rev. D} {\bf 61} 127304

\bibitem{Baiotti10} Baiotti L, Damour T,  Giacomazzo B,  Nagar A and  Rezzolla L 2010, {\it Phys. Rev. Lett.}  {\bf 105} 261101

\bibitem{Bose10} Bose, Sukanta; Ghosh, Shaon; Ajith, P. 2010 {\it Class. Quant. Grav} {\bf 27} 114001

\bibitem{Read13} Read J S, Baiotti L, Creighton J D E, Friedman J L., Giacomazzo B,  Kyutoku K, Markakis C, Rezzolla L, Shibata M and  Taniguchi K 2013, {\it Phys. Rev. D} {\bf 88} 044042

\bibitem{Maselli13} Maselli A, Gualtieri L and Ferrari, V 2013 {\it Phys. Rev. D} {\bf 88} 104040

\bibitem{Bauswein15a} Bauswein A and Stergioulas N 2015 {\it Phys. Rev. D} {\bf 91} 124056

\bibitem{Bauswein15b} Bauswein A and Stergioulas N and Janka, H-T 2015 {\it EPLA} in press

\bibitem{Fryer15} Fryer C L, Belczynski K, Ramirez-Ruiz E, Rosswog S, Shen, G and Steiner, A W 2015 {\it Astrophys. J.} {\bf 812} 24

\bibitem{Dietrich15} Dietrich T,  Moldenhauer N, Johnson-McDaniel N K, Bernuzzi S, Markakis C M, Br\"ugmann B and Tichy W 2015 {\it Phys. Rev. D} {\bf92}  124007

\bibitem{duez03} Duez M D, Marronetti P,  Shapiro S L and Baumgarte T W 2003 {\it Phys. Rev. D} {\bf 67} 024004

\bibitem{Marronetti03} Marronetti P,  Duez M D, Shapiro S L and Baumgarte T W 2003 {\it Phys. Rev. Lett.} {\bf 92} 141101

\bibitem{Miller04} Miller M,  Gressman P, and Suen W-M, 2004 {\it Phys. Rev. D} {\bf 69} 064026

\bibitem{Miller05} Miller M 2005 {\it Phys. Rev. D} {\bf 71} 104016

\bibitem{Miller07} Miller M 2007 {\it Phys. Rev. D} {\bf 75} 024001

\bibitem{Uryu06} Uryu K, Limousin F, Friedman J L, Gourgoulhon E and Shibata M 2006 {\it Phys. Rev. Lett.} {\bf 97} 171101

\bibitem{Kiuchi09} Kiuchi K, Sekiguchi Y, Shibata M and Taniguchi K 2009 {\it Phys. Rev. D} {\bf 80} 064037

\bibitem{Bernuzzi15}  Bernuzzi, S, Nagar, A, Dietrich, T and  Damour, T 2015 {\it Phys. Rev. Lett.} {\bf 114} 161103

\bibitem{DePietri15} De Pietri R, Feo A, Maione F and L\"offler F 2015 eprint arXiv:1509.08804 

\bibitem{Flanagan99} Flanagan, E E 1999, Phys. Rev. Lett., 82, 1354 

\bibitem{wmbook} Wilson J R and Mathews G J 2003 {\it Relativistic Numerical Hydrodynamics},
(Cambridge University Press, Cambridge, United Kingdom)

\bibitem{haywood} Haywood, J. R. 2006 PhD Thesis, University of Notre Dame

\bibitem{Gourgoulhon01} Gourgoulhon E, Grandcle'ment P, Taniguchi K, Marck J. and
Bonazzola S 2001 Phys. Rev. {\bf D 63} 064029 

\bibitem{Taniguchi01} Taniguchi K, Gourgoulhon E and Bonazzola S 2001  Phys. Rev. {\bf D64} 064012

\bibitem{Lan16} Lan N Q, Suh, I-S Mathews G J and Reese J H 2016 {\it Comm.. Phys.} In Press

\bibitem{mw97} Mathews G J and Wilson J R, 1997 {\it Astrophys. J.} {\bf 482} 929

\bibitem{adm} Arnowitt R, Deser S, and Misner C W 2008 {\it Gen. Rel. Grav.} {\bf 40} 1997

\bibitem{york79} York J W Jr 1979 {\it Sources of Gravitational Radiation} ed Smarr L L
(Cambridge University Press, Cambridge, UK) p. 83,

\bibitem{thorne80}  Thorne K S 1980 {\it Rev. Mod. Phys.} {\bf 52} 299

\bibitem{evansdiss} Evans C R 1984  {\it A Method for Numerical Relativity: Simulation of Axisymmetric
                    Gravitational Collapse and Gravitational Radiation Generation.}, PhD thesis,
                    University of Texas at Austin

\bibitem{flan99} Flanagan E E 1999 {\it Phys. Rev. Lett.} {\bf 82} 1354

\bibitem{wilson79} Wilson J R  1979 {\it Sources of Gravitational Radiation} ed Smarr L L
(Cambridge University Press, Cambridge, UK) p.  423

\bibitem{mtw} Misner C W, Thorne K S, and Wheeler J A 1973, {\it Gravitation} (Freeman, San Francisco)

\bibitem{maywil93} Mayle R W, Tavani M and Wilson J R, 1993 {\it Astrophys. J.} {\bf 418} 398

\bibitem{ls91}  Lattimer J Mand Swesty F D  1991 {\it Nucl. Phys. A} {\bf 535} 331

\bibitem{glen96} Glendenning N K  1996 {Compact stars, nuclear physics, particle physics, and general
relativity}, (Springer-Verlag,) New York

\bibitem{garg04}  Garg U 2004 {\it Nucl Phys. A} {\bf 731} 3

\bibitem{cut93} Cutler C, Apostolatos T A, Bildsten L, Finn L S, Flanagan E E, Kennefick D, 
Markovic D M, Ori A,  Poisson E and Sussman G J, 1993 \PRL {\bf 70} 2984

\bibitem{kidder} Kidder L E, Will C M  and Wiseman A G 1993 {\it Phys. Rev. D}  {\bf 47} 4183

 \bibitem{shibu01} Shibata M and Uryu K 2001 {\it Phys. Rev. D} {\bf 64} 104017

\end{thebibliography}
\end{document}